\documentclass[journal,10pt]{IEEEtran}
\usepackage{csquotes}
\usepackage{tikz}
\usepackage{pgfplots}
\usepackage{amsmath, amssymb}
\usetikzlibrary{plotmarks}
\usepgfplotslibrary{fillbetween}
\usetikzlibrary{arrows, decorations.markings, decorations}
\usetikzlibrary{arrows,arrows.meta,patterns,positioning,shapes}
\usetikzlibrary{plotmarks}
\usepgfplotslibrary{fillbetween}
\usetikzlibrary{arrows, decorations.markings, decorations}
\usepgflibrary{patterns.meta}
\usepackage{acro}
\usepackage{booktabs}
\usepackage{pgfplots}
\usepgfplotslibrary{groupplots,dateplot}
\usepackage[all]{nowidow}
\usepackage{soul}
\usepackage{tabularx,multirow}
\usepackage{xspace}
\usepackage{tikz}
\usepackage{bm}
\usetikzlibrary{arrows,arrows.meta,patterns,positioning,shapes}
\usepackage{soul}
\usepackage{enumitem}
\usepackage[ruled,vlined]{algorithm2e}
\usepackage{makecell}
\usepackage{siunitx}
\usepackage{multirow}
\usepackage{adjustbox}
\usepackage{csvsimple}
\usepackage[capitalise]{cleveref}
\usepackage{url}

\definecolor{ForestGreen}{RGB}{34,139,34}

\usepackage{todonotes}

\pgfplotsset{
    compat=1.15,
}

\definecolor{tabblue}{HTML}{5075b2}
\definecolor{tabred}{HTML}{bf575a}
\definecolor{tabgreen}{HTML}{71b582}
\definecolor{tabyellow}{HTML}{cdba77}
\definecolor{tabpink}{HTML}{DA70D6}
\DeclareSIUnit{\pp}{\textup{p.p.}}
\usepackage{acro}
\DeclareAcronym{fl}{short=FL,long=Federated Learning,short-indefinite=an}
\DeclareAcronym{iot}{short=IoT,long=internet of things,short-indefinite=an}
\DeclareAcronym{nn}{short=NN,long=neural network,short-indefinite=an}
\DeclareAcronym{cnn}{short=CNN, long=convolutional neural network}
\DeclareAcronym{fedavg}{short=FedAvg, long=Federated Averaging}
\DeclareAcronym{iid}{short=iid, long=independent and identically distributed, short-indefinite=an}
\DeclareAcronym{ml}{short=ML, long=machine learning}
\DeclareAcronym{sgd}{short=SGD, long=stochastic gradient descent}
\DeclareAcronym{nas}{short=NAS, long=neural architecture search}
\DeclareAcronym{flop}{short=FLOP, long=Floating Point Operation}
\DeclareAcronym{mac}{short=MAC, long=Multiply and Accumulate}
\DeclareAcronym{sa}{short=SA, long=Systolic Array}
\DeclareAcronym{simd}{short=SIMD, long=Single Instruction Multiple Data}
\DeclareAcronym{alu}{short=ALU, long=Arithmetic Logic Unit}
\DeclareAcronym{mbm}{short=MBM, long=Minimally Biased Multiplier}

\newcommand{\etal}{\emph{et al.}\xspace}

\ifCLASSINFOpdf
\else
\fi
\begin{document}

%
\title{Energy-Aware Heterogeneous Federated Learning via Approximate DNN Accelerators}
%
%
%

\author{Kilian~Pfeiffer,
        Konstantinos~Balaskas,
        Kostas~Siozios,~\IEEEmembership{Senior Member,~IEEE}
        and~Jörg~Henkel,~\IEEEmembership{Fellow,~IEEE}
\thanks{K. Pfeiffer, and J. Henkel are with the Chair for Embedded Systems, Karlsruhe Institute of Technology, 76131 Karlsruhe, Germany (e-mail  \{kilian.pfeiffer, henkel\}@kit.edu).}
\thanks{K. Balaskas is with Department of Computer Engineering and Informatics, University of Patras, 26504 Patras, Greece (e-mail  kompalas@ceid.upatras.gr).}
\thanks{K. Siozios is with Department of Physics, Aristotle University of Thessaloniki, 54124 Thessaloniki, Greece (e-mail  ksiop@auth.gr).}
\thanks{The authors would like to thank Lokesh Siddhu for his helpful discussions on the technical implementation. This work was partially funded by the German Research Foundation (DFG) as part of project NA3OS (grant: 524986327). The authors acknowledge support by the state of Baden-Württemberg through bwHPC.}
}
\maketitle

\makeatletter
\def\ps@IEEEtitlepagestyle{
  \def\@oddfoot{\mycopyrightnotice}
  \def\@evenfoot{}
}
\def\mycopyrightnotice{
  {\footnotesize
  \begin{minipage}{\textwidth}
  \centering
  © 2024 IEEE. Personal use of this material is permitted. Permission from IEEE must be
  obtained for all other uses, in any current or future media, including
  reprinting/republishing this material for advertising or promotional purposes, creating new
  collective works, for resale or redistribution to servers or lists, or reuse of any copyrighted
  component of this work in other works.
  \end{minipage}
  }
}
\begin{abstract}
In Federated Learning (FL), devices that participate in the training usually have heterogeneous resources, i.e., energy availability.
In current deployments of FL, devices that do not fulfill certain hardware requirements are often dropped from the collaborative training.
However, dropping devices in FL can degrade training accuracy and introduce bias or unfairness.
Several works have tackled this problem on an algorithm level, e.g., by letting constrained devices train a subset of the server neural network (NN) model.
However, it has been observed that these techniques are not effective w.r.t. accuracy.
Importantly, they make simplistic assumptions about devices' resources via indirect metrics such as multiply accumulate (MAC) operations or peak memory requirements. We observe that memory access costs (that are currently not considered in simplistic metrics) have a significant impact on the energy consumption.
In this work, for the first time, we consider on-device accelerator design for FL with heterogeneous devices.
We utilize compressed arithmetic formats and approximate computing, targeting to satisfy limited energy budgets.
Using a hardware-aware energy model, we observe that, contrary to the state of the art's moderate energy reduction, our technique allows for lowering the energy requirements (by $4\times$) while maintaining higher accuracy.
\end{abstract}


%
\IEEEpeerreviewmaketitle
\noindent
\begin{minipage}{\columnwidth}  
  \footnotesize
  \phantom{000}This article has been accepted for publication in IEEE Transactions on Computer-Aided Design of Integrated Circuits and Systems (TCAD). This is the author's version which has not been fully edited and content may change prior to final publication. Digital Object Identifier: 10.1109/TCAD.2024.3509793.
\end{minipage}

\section{Introduction}
\label{sec:introduction}
Deep learning has achieved impressive results in many domains in recent years. However, these breakthroughs are fueled by massive amounts of centrally available data. After centralized training, an inference-optimized model is often deployed on embedded systems, such as smartphones or \ac{iot} devices. However, data that resides on such devices is often privacy-sensitive, and must not be uploaded centrally to a server. \Ac{fl} has emerged as an alternative to centralized training. In \ac{fl}, the \textit{server} is responsible for aggregating the \ac{nn} models, while training is performed locally with the data that resides on the devices. Training of \ac{nn} models, however, remains an energy-intensive task: while in the centralized case, training can be done on power-hungry GPUs that consume several hundreds of watts, training on embedded devices is limited to a few watts at most. Additionally, devices that participate in \ac{fl} usually have \textit{heterogeneous} energy availability for training~\cite{pfeiffer2023survey}.

In current deployments of \ac{fl}, such as Google GBoard~\cite{yang2018applied}, devices that do not fulfill certain hardware requirements, or are not connected to wall power, are dropped from the training.
While these approaches may simplify device management, they have significant drawbacks.
Specifically,
dropping devices limits the available training data, and 
as a result, the global model may fail to generalize well across all data distributions.
Hence, dropping devices can lower the accuracy of the trained \ac{nn} model.
This can be especially critical as energy-constrained devices often hold important, unique data--often inaccessible by other devices--which cannot contribute to the training process, thus leading to bias and unfairness~\cite{pfeiffer2023survey,maeng2022towards}.
This necessitates to include them in the training.

\Ac{fl} training with heterogeneous devices has already been approached by several algorithm-level techniques, that aim to reduce the number of operations for constrained devices at run time. A main branch of works, such as HeteroFL~\cite{diao2020heterofl} (and~\cite{horvath2021fjord, alam2022fedrolex, caldas2018expanding}) employ subsets of the server \ac{nn} to energy-constrained devices.
However, it has been shown that they are inefficient and the aggregation of subsets can even harm the \ac{nn} model performance~\cite{cheng2022does, pfeiffer2023cocofl}.
Importantly, all these works make simplistic assumptions about the devices' resources via indirect metrics (e.g., \ac{mac} operations, \acp{flop}, and peak memory requirements),
which do not accurately reflect the actual energy cost, especially considering memory accesses.
We show in our evaluation (see \cref{sec:experimental_evaluation}), that relying on these assumptions leads to vastly different energy estimations when using a more realistic training energy model.

\begin{figure}
    \centering
    \includegraphics[page=1]{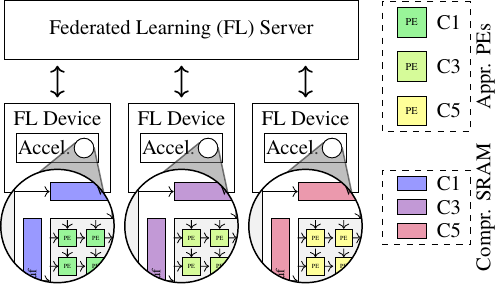} 
    \caption{We envision \iac{fl} system, where each \ac{fl} device is equipped with a specific systolic array accelerator, where its specifications are decided at design time. Depending on the devices' environments, accelerators with specific approximate processing elements and compressed SRAM are used. The full details of our accelerator design are visualized in \cref{fig:accelerator}. The compression and approximation levels C1-C5 are listed in \cref{tab:accelerator}.}
    \label{fig:flsystem}
\end{figure}

In this work, we introduce a third, promising, and unexplored direction besides dropping devices and training model subsets: hardware accelerator design for FL environments. Specifically, we explore how the heterogeneity challenges in \ac{fl} can be addressed already at the design stage.
Our \ac{fl} system is visualized in \cref{fig:flsystem}.
We assume that it is a priori known that some devices have tighter energy constraints (e.g., they may have to rely on energy harvesting, or have different cooling capabilities).
In our work, we purposefully exploit this known device heterogeneity by designing training-capable \ac{nn} accelerators to satisfy the constraints.
Domain-specific \ac{nn} accelerators have emerged as the ubiquitous solution for reducing the immense energy consumption (compared to CPU, GPU) of \acp{nn} and thus enabling their deployment on energy-constrained devices~\cite{Chen:SSC:2017:eyeriss}.

Furthermore, we integrate approximate computing principles within our accelerator designs.
Hardware approximation techniques can exploit the inherent error resilience of \ac{nn} training and drastically reduce the (considerable) energy demands of \ac{nn} accelerators~\cite{Armen:ACMSurv:2023:ac}.
Approximate computing, though an established design paradigm, introduces new opportunities within a realistic distributed environment of \ac{fl}, such as maintaining low device energy consumption whilst regulating manufacturing costs.
Consider a scenario where, compared to cutting-edge fabrication technology, certain devices (e.g., smartphones) are manufactured at cheaper technology nodes, for cost-reduction purposes.
In order to maintain the energy available for training at similar levels across devices, approximate computing can be orthogonally applied on cheaper devices to compensate for their (inevitably) higher energy consumption.
Disregarding important design-time energy information of \ac{fl} devices (with the use of algorithm-level estimations for example) may cause failure to comply with such realistic constraints (e.g., fabrication cost), which could in turn result in devices being dropped from training.
We envision our design-time approach as particularly beneficial in domains such as healthcare applications (e.g., healthcare wearables), where low energy consumption and high data privacy protection are required~\cite{baghersalimi2021personalized:r2_rebuttal, baghersalimi2023decentralized:r3_rebuttal}.

Our accelerators follow the architecture of widely adopted \ac{mac}-based inference accelerators with the use of \acp{sa} (e.g., TPU~\cite{tpu_v1}), but enhanced to support training capabilities.
Importantly, they are equipped with a \ac{simd} array, such that convolutional, non-convolutional, and gradient operations are gracefully handled.
By design, each accelerator supports a selected (compressed) arithmetic format in both storage and computation, to alleviate the overhead of redundant FP32 formats.
We further increase the energy efficiency of highly constrained devices by employing approximate floating-point multipliers, via approximate mantissa multiplications.
The use of approximate computing, via the combination of approximate computations and compressed storage arithmetic, facilitates the creation of diverse accelerator designs with a wide range of energy/accuracy trade-offs, and as an extension, the satisfaction of energy constraints of heterogeneous \ac{fl} devices.
Each accelerator's energy consumption is reliably quantified with an energy model, accounting for both computational and storage costs of all training operations.
Our accelerators significantly reduce energy consumption for constrained devices while maintaining the ability to perform full model training, thus improving both efficiency and fairness in \ac{fl} by enabling devices to participate meaningfully in the training process.
We showcase in our evaluation that, while state-of-the-art approaches can only moderately scale the energy consumption,
our design-time techniques are effective in reducing the energy for training (without downscaling the \ac{nn} model), while allowing constrained devices to make an impact on the global \ac{fl} model, thus maintaining model fairness.

In summary, we make the following novel contributions:
\begin{itemize}
    \item To the best of our knowledge, we are the first to consider hardware design for \ac{fl} with heterogeneous devices. 
    \item We design training-capable \ac{nn} accelerators to meet the energy targets of an \ac{fl} environment.
    Our accelerators utilize compressed arithmetic formats and approximate computing for higher energy efficiency.
    \item We reliably quantify the energy consumption of our devices during training with a storage- and computation-aware energy model, instead of relying on indirect metrics. 
    \item
    We show in our evaluation that, contrary to the state-of-the-art's suboptimal energy reduction, our technique can lower the energy requirements up to 4$\times$, whilst accuracy, and therefore fairness for constrained devices are maintained.
\end{itemize}

\section{FL Problem Statement}
\label{sec:problem_statement}
For our \ac{fl} setting, we consider \ac{fedavg}~\cite{mcmahan2017communicationefficient}, which comprises a single \textit{server} and many participating \textit{devices} $c \in \mathcal{C}$ that act as clients. Each device has its own local dataset $\mathcal D_c$ consisting of data $\bm X$ and labels $y$. \Ac{fl} is usually done in synchronous rounds~$r \in [1,\ldots, R]$. In each round, a subset of devices $\mathcal{C}^{r}$ pulls the latest \ac{nn} parameters from the server and performs \ac{sgd}-based training with its local data. After a limited amount of epochs, each device uploads its locally trained parameters $w_c^{r}$ to the server. After each round, the server averages the local models by using
\begin{equation}
    w^{r+1} = \frac{1}{\sum_{c\in C^{(r)}|D_c|}}\sum_{c \in \mathcal{C}^{(r)}} |\mathcal D_c| w_c^{r},
\end{equation}
and redistributes the new averaged model $w^{r+1}$ to the devices for training in the subsequent round. Further, we assume that the devices deployed for \ac{fl} are subject to heterogeneous energy budgets, and their budgets are known at design time.

To enable a fair comparison to state-of-the-art techniques in \cref{sec:experimental_evaluation}, we assume that all accelerators are manufactured with the same technology node and compare them based on energy constraints.

Our goal is to maximize the accuracy of the global \ac{fl} model after~$R$ training rounds, while respecting energy constraints of a group of devices. Additionally, if data is \emph{resource-correlated (rc)}~\cite{maeng2022towards} non-\ac{iid}, we want to ensure that constrained devices can make a meaningful contribution to the global model, thus ensuring fairness (details are given in~\cref{sec:experimental_setup}).
\section{Related Work}
\label{sec:related_work}
\textbf{Computational heterogeneity in \ac{fl}:}
Computational heterogeneity in \ac{fl} has gained some attention in recent years. Caldas~\etal~\cite{caldas2018expanding} were the first to apply training of lower complexity models in \ac{fl}. Specifically, they randomly select convolutional filters of a larger server model, and randomly per device, per \ac{fl} round, select a subset of all filters, repack the selected filters to a dense matrix, and only train the submodel on the devices. However, they introduced no heterogeneity mechanism. 
This is addressed by HeteroFL~\cite{diao2020heterofl} (similarly~\cite{horvath2021fjord}).
In these works, constrained devices always train the same subset of the full model.
Depending on their capabilities, devices train a larger portion of the server model.
In FedRolex~\cite{alam2022fedrolex} a rolling window scheme is presented, where the subset is shifted over the full server model on a per-round basis.
Beyond subsets, also other mechanisms have been proposed.
In CoCoFL~\cite{pfeiffer2023cocofl} a quantization and freezing scheme is presented, where layers are frozen and calculated with integer precision, however, the technique relies on training on CPUs.
FedProx~\cite{li2020federated} tolerates partial work (e.g., fewer mini-batches) from constrained devices and uses a norm that enforces local parameters to remain close to the server parameters.
FedHM~\cite{yao2021fedhm} uses low-rank factorization to create lower-complexity submodels for constrained devices.
Lastly, works like ZeroFL~\cite{qiu2022zerofl} utilize sparse matrices for training and control the computational complexity through the sparsity level. Recently, in addition to reducing model complexity,  clustering and data-level based techniques have also been considered~\cite{ye2023heterogeneous}.

\textit{In summary, the majority of these works lower the \ac{nn} complexity (i.e., by reducing the number of operations) at run time, but do not address heterogeneity at design time. However, while operations are reduced, it has been shown that using \ac{nn} subsets can be inefficient w.r.t. accuracy~\cite{cheng2022does}.}

\textbf{Training energy estimation in heterogeneous \ac{fl}:} Works on energy estimation in \ac{fl} primarily focus on modeling the energy required for communication. Tran~\etal~\cite{tran2019federated} study energy trade-offs between communication and computation. However, their energy model for computation only considers CPU cycles and frequency (assuming training is done on a CPU), without considering memory. Works that study heterogeneity~\cite{caldas1812leaf, diao2020heterofl, horvath2021fjord, alam2022fedrolex} model resource requirements for training only by using the number of parameters. In CoCoFL~\cite{pfeiffer2023cocofl}, training time and peak memory are considered.

\textit{In summary, all these works model heterogeneity by using abstract metrics like \acp{mac}, \acp{flop}, number of \ac{nn} parameters, or peak memory,
but do not directly (or reliably) model energy to ensure energy constraints are satisfied.}

\textbf{Accelerators for \ac{nn} Training}:
Focusing on \ac{nn} hardware acceleration, on-device training has been the focus of several works in recent years~\cite{9774739}.
Dataflow optimizations have been proposed for efficient computations of all three training stages (i.e., forward pass, gradient calculation, weight update).
In~\cite{yang_procrustes_2020}, this is accompanied by custom accelerator microarchitectures and load-balancing techniques.
Authors in~\cite{awad_fpraker_2021} discard outlier values from a power-of-two numerical conversion data range, such that operations are accelerated.
Enabling sparse on-device training has shown significant interest~\cite{qin_sigma_2020, nakahara_fpga-based_2019}, but typically involves redundant floating-point precision (i.e., FP32), such that accuracy is retained.
Contrarily, lower precision prototypes~\cite{Lee:ISSCL:2019:mixedp, choi_energy-efficient_2020, venkataramani_rapid_2021} employ fixed-point or integer arithmetic but require significant profiling to identify quantization-resilient operations.
Hybrid solutions (e.g., GPU coupled with FPGAs) have also been introduced~\cite{he_enabling_2021} to tackle workload diversity within \ac{nn} training.
In-memory computing figures prominently among current research trends for on-device training~\cite{imani_floatpim_2019, jiang_two-way_2020}, whilst questions about stability remain yet unanswered.
In~\cite{esmaeilzadeh2023performance}, a flexible accelerator coupled with comprehensive analytical models that target all training operations is proposed. 
In \cite{yang2020fpga}, FPGA acceleration of homomorphic encryption is used in conjunction with \ac{fl}.
None of these works consider coordinated distributed training, i.e., applicability to \ac{fl}.

\textit{In summary, the works above utilize accelerators to enable on-device training, but mostly redundant floating-point formats are considered, and support for non-convolutional/GEMM operations is insufficient.}
\section{Accelerator Design for Heterogeneous Devices in FL}
\label{sec:technique}
\begin{figure}
    \centering
    \includegraphics[page=1]{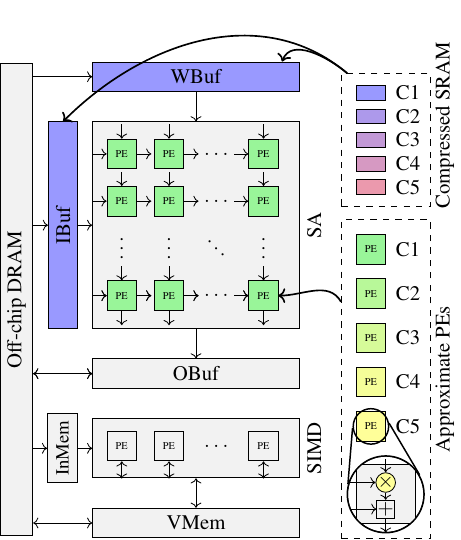} 
    \caption{Schematic overview of our designed training-capable accelerator, comprising of a \ac{sa}, \ac{simd} array, on-chip SRAM buffers and off-chip DRAM. \ac{mac}-based PEs (light green) within the \ac{sa} contain an approximate mantissa multiplier, which dictates the compressed format for SRAM storage (purple). Gray components remain in FP32 format.}
    \label{fig:accelerator}
\end{figure}
We aim, for the first time, to exploit device heterogeneity in an \ac{fl} environment at design time, such that the energy consumption of each device is accurately quantified and thus satisfies its inherent energy budget.
To that end, we design a set of  ASIC-based training-capable \ac{nn} accelerators.
Each accelerator is flexible enough to handle all necessary training operations (including non-convolutional operations and gradient calculations), but specialized to meet the encompassing device's energy budget via compressed arithmetic formats and hardware approximation.
Importantly, in contrast to the state-of-the-art, each device trains with the same (i.e., not downscaled) \ac{nn} architecture, but uses different approximate accelerators to account for its unique energy budget.
To quantify the energy gains through approximation, we introduce an energy model that considers both the computational and memory costs without relying on simplistic hardware metrics (e.g., \acp{flop}).

\subsection{Accelerator design}
\label{sec:accelerator_design}
Our proposed accelerator design is presented in \cref{fig:accelerator}, similarly to~\cite{esmaeilzadeh2023performance}.
Floating-point arithmetic is used to execute all training operations.
Convolutional operations are handled by an \ac{sa}, supporting weight-stationary dataflow.
The \ac{sa} is complemented by on-chip SRAM buffers (namely, IBuf, WBuf, and OBuf in \cref{fig:accelerator}) which store the necessary data (inputs, weights, and outputs, respectively).
PEs encapsulate a single \ac{mac} unit and two register files, for temporary storage (and propagation) of input activations and partial sums, per systolic guidelines.
Non-convolutional (e.g., BatchNorm) and gradient operations are facilitated by a 1-D \ac{simd} array, comprising parallel \ac{alu}-based \ac{simd} cores.
SRAM buffers are responsible for storing the \ac{alu} instructions (InMem) and data (VMem).
Each buffer in our accelerator communicates directly with the off-chip DRAM (DDR4).

The flexibility of our architecture, combining \ac{sa} and \ac{simd}, allows for the execution of all \ac{nn} operations on-chip. This capability is not supported by typical inference accelerators, such as TPU-V1~\cite{tpu_v1}. Nevertheless, our training-capable accelerators inherit several features from existing systolic architectures (e.g., weight-stationary dataflow, 2D \ac{sa}). 

Importantly, our energy model (see \cref{subsec:energy_model}) is agnostic to the specific accelerator architecture. Thus, existing training accelerators (e.g.,~\cite{esmaeilzadeh2023performance}) can be integrated into our framework, provided they support on-chip execution of all NN operations.

Lastly, we emphasize that the use of floating-point arithmetic is essential for supporting the wide dynamic range of training parameters and maintaining accuracy. This contrasts with inference, where fixed-point or low-bit integer arithmetic is preferred and is not the focus of our work. To address the redundancy inherent in FP32, our ASIC accelerators employ targeted approximations on both memory and computations, as detailed in the following section.
Note that while our approach primarily focuses on reducing the energy budget for \ac{fl} training, its flexible design, capable of accelerating both forward and backward passes, naturally extends to inference tasks, which are computationally less demanding. Given the effectiveness of approximate computing in reducing energy consumption during inference, we are confident that our hardware can seamlessly support inference with similar benefits in terms of accuracy and energy efficiency.

\subsection{Hardware Approximations}
\label{sec:hardware_approximations}
During \ac{fl} training, the major energy bottleneck for resource-constrained devices is the massive number of \ac{mac} operations that need to be performed between inputs and trained weights. We therefore target the \ac{mac} unit of our accelerators with hardware approximations to reduce its energy requirements, thus enabling devices with even the tightest energy budgets to participate meaningfully in training.

\begin{figure}
    \centering
    \includegraphics[page=1]{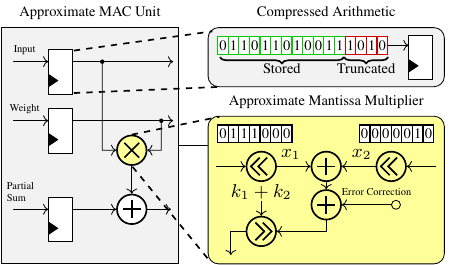} 
    \caption{Schematic overview of our proposed approximate \ac{mac} unit (left). Each \ac{mac} unit in our \ac{sa} is equipped with hardware approximation techniques at both the memory and computational levels (right). We utilize compressed arithmetic formats to fetch and store weights and input activations. Additionally, we employ the state-of-the-art \ac{mbm} floating-point multiplier~\cite{saadat2018minimally}, which leverages a linear version of the logarithmic multiplication property along with error correction for approximate mantissa multiplications.}
    \label{fig:mac_unit}
\end{figure}

Our comprising \ac{mac} units support floating-point arithmetic and can be designed as either exact or approximate.
A schematic overview of our proposed approximate \ac{mac} units--along with the introduced hardware approximations--can be seen in \cref{fig:mac_unit}.
Approximations are introduced with the use of the state-of-the-art approximate \ac{mbm}~\cite{saadat2018minimally} as the mantissa multiplier.
Note, our technique is not reliant on specific approximate circuits, thus any other floating-point multiplier could be seamlessly supported (e.g.,~\cite{chen2021pam, li2020accuracy}).
\ac{mbm} relies on a linear version of the logarithmic multiplication property for approximate mantissa multiplication.
In detail, \ac{mbm} is built upon Mitchell's linear approximation, which approximates the binary log and antilog of operands (mantissas) by determining the leading one (the characteristic) and treating the remaining bits as the fractional part. The logs of two operands are added, and the reverse process is used to obtain the approximate product, thereby reducing computational complexity. Assuming two unsigned integer multiplicands (mantissas) of~$N$ bits, with characteristics~$x_1, x_2$ and leading-one positions at~$k_1, k_2$, the approximate product~$\tilde{P}$ generated by the \ac{mbm} multiplier is as follows:
\begin{equation}
    \tilde{P} =
    \begin{cases}
        2^{k_1 + k_2} ( 1 + x_1 + x_2 + c) & \text{if}\, x_1 + x_2 < 1 \\
        2^{k_1 + k_2 + 1} ( x_1 + x_2 + \frac{c}{2})  & \text{if}\, x_1 + x_2 \ge 1,
    \end{cases}
\label{eq:mbm_log_property}
\end{equation}
where~$c$ is an error correction term derived from studying the error magnitude curve between unsigned integers. We refer the reader to the original work in~\cite{saadat2018minimally} for further information.

Interestingly, as seen in \cref{fig:mac_unit} and deduced by \cref{eq:mbm_log_property}, our approximate mantissa multiplier is, in fact, multiplier-less, since multipliers are substituted by cost-effective adders, and shift operations are used. This significantly reduces the energy consumption of our accelerators. The hardware approximation technique simplifies computations of the mantissa product, allowing for a more efficient multiplier implementation. Furthermore, it is configured to truncate operand LSBs to the desired degree. Targeting the mantissa multiplication with precision-related approximation helps reduce the energy consumption of the floating-point multiplier, while simultaneously introducing minimal errors.

Multipliers represent the most power-hungry component of a \ac{mac} unit, so approximations applied to the multiplication operation have a significant effect on energy reduction. Additionally, within a floating-point multiplier, mantissa multiplication consumes more power and has less impact on multiplication accuracy (due to lower bit significance) than exponent addition~\cite{saadat2018minimally}, making it an ideal candidate for approximation. On the other hand, the adder is responsible for accumulating partial sums, contributing more directly to final output activations. As it has a more direct impact on output accuracy, the adder is designed to be exact.

In order to reduce the elevated energy consumption stemming from redundant FP32 data, we design \ac{mac} units and SRAM buffers to support compressed arithmetic formats.
Compression in our case refers to reduced mantissa precision within the floating-point format.
For example, assuming bfloat16 arithmetic as a baseline (i.e., 7 mantissa bits), if \ac{mbm} truncates $2$ mantissa bits for computations, we store and propagate only the $5$ unmasked MSBs to on-chip memory.
In practice, we configure the SRAM bus width to efficiently transfer batches of such compressed floating-point data.
The proposed techniques serve as evidence of approximate computing's effectiveness on energy reduction of accelerators, orthogonally to the \ac{fl} setting.
Thus, any other circuit-level approximation (e.g., netlist pruning) is expected to behave similarly.

Note, that our design strategy mainly focuses on reducing the energy consumption of the \ac{sa}, i.e. the cost of \ac{mac} operations and the relevant SRAM energy accesses.
As the costly convolutional operations are allocated to the \ac{sa}, it makes up for the majority of dynamic energy consumed in the entire accelerator (more details in Section~\ref{sec:experimental_evaluation} and \cref{fig:energy_components}).
Static energy during idleness of the unused component (\ac{sa}/\ac{simd}) is insignificant and omitted from our calculations.
For example, static/leakage power dissipated by modern \ac{mac} units consists~$<1\%$ of its dynamic power~\cite{Chhajed:CSSP:2022:bitmac}.
With the flexibility of utilizing compressed arithmetic formats (in both computational and SRAM memory levels), along with several degrees of approximation, our accelerators are capable of offering a wide range of energy/accuracy trade-offs, based on the expected energy budgets of \ac{fl} devices.

Overall, our approximations address both major components driving energy consumption: computations and memory transfers. Part of our approach involves quantization-based techniques. Compressed arithmetic formats entail trimming input data and weights. Our \ac{mac} units feature approximate mantissa multipliers, which leverage the approximate logarithmic multiplication property alongside precision scaling. This combination effectively reduces the energy consumption of heterogeneous \ac{fl} devices while achieving higher accuracy compared to state-of-the-art methods (see \cref{sec:experimental_evaluation}).

Further approximations beyond quantization, such as voltage overscaling, are orthogonal to our work but could be applied to our accelerators for additional energy savings. However, these methods would likely result in reduced training accuracy.

\subsection{Mapping Training Operations}
\label{sec:training_operations}
Here, we describe in detail the way each training operation is mapped to our accelerators.
Training consists of three distinct phases:
forward pass, gradient calculation (or backward pass), and weight update.
Outputs calculated during the forward pass contribute to the evaluation of the loss function, whose gradient w.r.t. each parameter is propagated in the opposite directions, thus constructing gradient matrices.
Gradients are consequently used to update parameters such that the error is minimized.
Although interconnected, each step demands different handling in terms of dataflow and mapping to the accelerator~\cite{lee_overview_2021}.
Note, that we refer to convolutional layers, since the rest are straightforwardly mapped to the \ac{simd} during each training phase.
Fully-connected layers follow similar steps, with the substitution of transposing instead of rotating weights, and point-wise multiplication instead of convolution, for the backward pass.

\textbf{Forward pass:}
Assuming a weight stationary dataflow (which is the case in our work), weights are directly mapped to PEs of the \ac{sa} unaltered.
Contrarily, input feature maps are tiled and streamed (horizontally broadcasted) in a row or column-wise fashion, performing the respective convolutions~\cite{yang_procrustes_2020}.
Each partial sum is stored and accumulated in the register file, and propagated (vertically reduced) to its systolic neighbors, thus forming output feature maps within the output buffer.

\textbf{Gradient calculation:}
For the backward pass (i.e., convolving weights with loss gradients w.r.t. output activations) to be executed, weights have to be re-ordered before being mapped to PEs.
Weight tensors are intra-channel-wise rotated and inter-channel-wise transposed, before being loaded onto the \ac{sa}.
Then, convolutional operations are carried out similarly to the forward pass, except that loss gradients from previous layers act as the streaming input.
Accumulation in this case produces the gradients w.r.t. inputs, which propagate to consequent layers.

\textbf{Weight update:}
Similar to the forward pass, inputs are horizontally broadcasted onto the \ac{sa}, to be then convolved with the loss gradients w.r.t. output activations.
The latter serves as the vertically reduced parameter, faithfully complying with weight-stationary rules.
Therefore, the formed gradients w.r.t. weights are unicasted and moved to the weight buffer, once the convolution terminates.
Note, that no tensor transformations are required during this step.

\subsection{Energy Model}
\label{subsec:energy_model}
In this section, our method for estimating the energy consumption of our training-capable accelerators is described.
For each \ac{fl}-capable device, we segment its total energy consumption into two distinct terms: computation-related and memory-related energy consumption ($E_\text{comp}$ and $E_\text{mem}$, respectively).
Given the architectural components of each accelerator for both computation (exact or approximate) and memory (see Section~\ref{subsec:accelerator_setup}), each term can be broken down into component-specific ones
\begin{flalign}
    \label{eq:total_energy}
    E_\text{total} = E_\text{mem} + E_\text{comp} &= \left( E_\text{DRAM} + E_\text{SRAM} \right) + \\
    &+ \left( E_\text{SA} + E_\text{SIMD} \right), \notag
\end{flalign}
where $E_\text{SA}$, $E_\text{SIMD}$ and $E_\text{DRAM}$ denote the energy consumption of the \ac{sa}, \ac{simd} array, and off-chip DRAM, respectively. The SRAM energy consumption is the energy consumption of each on-chip SRAM buffer, where $\mathcal{S}_\text{buf}$ represents a set of the buffers depicted in \cref{fig:accelerator} (i.e., IBuf, WBuf, OBuf, InMem and VMem).
We abstract the energy consumption of each buffer~$i \in \mathcal{S}_\text{buf}$ as the product of the number of accesses (read/write) ($\#A_i$) and the per-access energy cost ($e_i$), s.t.:
\begin{align}
    E_\text{mem} &= \#A_\text{DRAM} \times e_\text{DRAM} \; + \; \sum_{i \in \mathcal{S}_\text{buf}} \#A_{i} \times e_{i}.
\end{align}
The computational energy term in~\eqref{eq:total_energy} is simplified into a sum of products: the number of operations ($\#OP$) of each computational unit $i$ is multiplied by the per-operation energy cost ($c_i$):
\begin{align}
    E_\text{comp} = \#OP_\text{SIMD} \times c_\text{SIMD} \; + \; \#OP_\text{SA} \times c_\text{SA}. 
\end{align}

The above formulation of our energy model allows for fast and sufficiently accurate energy estimations during \ac{fl} training and does not rely on simplistic metrics to quantify the energy consumption.
Additionally, it is orthogonal to the underlying nature of computation (i.e., exact or approximate) and arithmetic format, and thus can easily scale to a wide range of heterogeneous accelerators.

\section{Experimental Setup}
\label{sec:experimental_setup}

\subsection{Hardware implementation}
\label{subsec:accelerator_setup}
Here, we describe the implementation details of our \ac{nn} accelerators and energy model.
Accelerators are equipped with a $16\times16$ \ac{mac}-based \ac{sa} and a 1-D \ac{simd} array of $16$ \ac{simd} parallel cores.
Off-chip DRAM size is set at \SI{2}{\giga\byte} and SRAM buffer sizes at \SI{64}{\kilo\byte}.
The SRAM bus width is determined by the compressed arithmetic format and is set at either $64$ or $60$ bits.

The implementation of our energy model is based on industry-strength tools and simulators.
Analytical models described in Genesys/SimDIT~\cite{esmaeilzadeh2023performance} are used to extract the number of accesses and operations.
Per-operation energy costs derive from synthesized computational units (i.e., \ac{mac} for the \ac{sa} and \ac{alu} for \ac{simd}), where the approximations and arithmetic format are embedded in the synthesizable RTL.
\ac{mac} units and \acp{alu} are synthesized with Synopsys Design Compiler (\texttt{compile\_ultra} command), using optimized arithmetic components from the commercial  Synopsys DesignWare library.
Synthesized designs are mapped to the NanGate \SI{45}{\nano\meter}\footnote{\url{https://si2.org/open-cell-library}} library.
Switching activity (via gate-level simulations) and average power consumption are obtained via QuestaSim and Synopsys PrimeTime, respectively.
CACTI~\cite{cacti} is used for SRAM/DRAM access energy estimations.

Our core energy measurements (see Section~\ref{subsec:energy_model}) can be found in \cref{tab:energy}.
Per-bit energy accurately reflects our approximation-induced benefits, since solely contiguous data are stored in the accelerator's on-chip memories.
Note, SRAMs of $60$ bus width have higher per-bit energy but ultimately pose a more efficient solution due to fewer bits to transfer (i.e., linked with compressed arithmetic formats).
\begin{table}[t]
    \small
    \setlength{\tabcolsep}{10pt}
    \renewcommand{\arraystretch}{1.1}
    \caption{Energy measurements for each component of our NN accelerator architectures, used within our energy model. We report the per-operation energy consumption of MAC/ALU, and the per-bit access energy for SRAM/DRAM.}
    \centering
    \begin{tabular}{l | c}
        \toprule
        \textbf{Component}          &   \textbf{Energy [\si{\pico\joule}]}
        \\ \midrule
        FP32 MAC unit (exact)       &   $26.8$
        \\ 
        bfloat16 MAC unit (exact)   &   $5.35$
        \\ 
        bfloat16 MAC unit (MBM-7)    &   $3.11$
        \\ 
        bfloat12 MAC unit (MBM-3)    &   $2.78$
        \\ 
        bfloat10 MAC unit (MBM-1)    &   $2.65$
        \\ 
        FP32 ALU                    &   $31.4$
        \\ 
        SRAM (64b bus width)        &  $0.401$
        \\ 
        SRAM (60b bus width)        &  $0.412$
        \\ 
        DRAM                        &  $41.0$
        \\
        \bottomrule
    \end{tabular}
    \label{tab:energy}
\end{table}
\begin{table}[t]
    \small
    \setlength{\tabcolsep}{4pt}
    \renewcommand{\arraystretch}{1.1}
    \caption{Accelerator designs (C1-C5) to target diverse heterogeneity scenarios. The compressed arithmetic format and the corresponding approximate mantissa multiplier are reported. Also, different \ac{nn} model scaling configurations (S1-S4), used by the state-of-the-art, are included.}
    \centering
    \begin{tabular}{l | c c c}
        \toprule
        \textbf{\#} & \textbf{SRAM format} & \textbf{MAC format} & \textbf{Approximation}
        \\ \midrule
         C1 & FP32 & FP32 & - 
         \\
         C2 & bfloat16 & bfloat16 & -
         \\
         C3 & bfloat16 & bfloat16 & MBM-7
         \\
         C4 & bfloat12 & bfloat12 & MBM-3
         \\
         C5 & bfloat10 & bfloat10 & MBM-1 \\
         \midrule
         S1-S4 & FP32 & FP32 & - \\
         \bottomrule
    \end{tabular}
    \label{tab:accelerator}
\end{table}

\begin{figure}
\centering
\includegraphics[page=1]{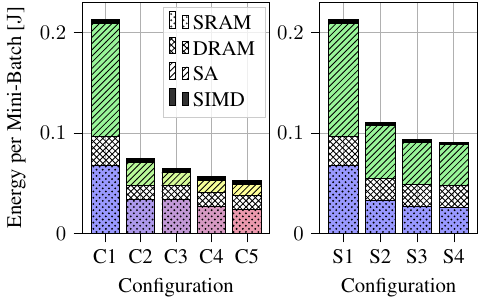} 
    \caption{Per-component energy breakdown for each accelerator configuration (C1-C5, left) and \ac{nn} subset (S1-S4, right), according to our energy model for a single mini-batch training step with ResNet20 and $3\times32\times32$ input size. SRAM accesses and \ac{sa} computations dominate the overall energy consumption. Color coding according to \cref{fig:accelerator}.}
    \label{fig:energy_components}
\end{figure}

\textbf{Accelerator heterogeneity:}
To target diverse heterogeneity scenarios, we evaluate using several accelerator configurations (namely, C1-C5, presented in \cref{tab:accelerator}), comprising various arithmetic formats and approximations.
Notice, C1-C5 can be seen in both \cref{fig:flsystem} and \cref{fig:accelerator}.
C1 (i.e., computations and on-chip storage are conducted in FP32) serves as our baseline, representing devices without energy constraints.
Similarly, C2 uses the bfloat16 arithmetic format, along with exact computations (i.e., accurate mantissa multiplications).
In C3-C5, the mantissa multiplier of the \ac{mac} unit is replaced by a state-of-the-art \ac{mbm} (\ac{mbm}-X denotes $\text{X}$ truncated mantissa LSBs), facilitating compressed arithmetic formats for storage (bfloatX uses $\text{X}<16$ bits, including $1$ sign bit, $8$ exponent bits and $\text{X}-9$ mantissa bits).
We map bfloat16 to SRAMs of $64$ bus width and bfloat12/10 to $60$ for efficient memory transfers.
\cref{fig:energy_components}(left) reveals a per-component energy breakdown for each accelerator configuration with ResNet20, mini-batch size of 32 and input size of $3\times32 \times32$.
It is observed that both \ac{sa} and SRAM constitute the energy bottlenecks for our accelerators, which motivates us to focus on these components with energy reduction mechanisms. Memory accesses are a significant contributor to training energy, with SRAM and DRAM accesses accounting for $31.7\%$ and $13.3\%$ of the total energy, respectively. This underscores the importance of incorporating memory accesses into the energy model, rather than relying solely on \ac{mac} costs. Finally, the progressive increase in approximation (and compression) from C1 to C5 leads to a corresponding reduction in energy consumption, enabling the adoption of even tighter energy constraints.

\subsection{\ac{fl} setup}
\label{subsec:fl_setup}
We evaluate our technique in \iac{fl} setting (described in Section~\ref{sec:problem_statement}) on the task of image classification, with the use of the CIFAR10, CIFAR100~\cite{krizhevsky2009learning}, TinyImageNet~\cite{le2015tiny} and FEMNIST datasets, obtained from the \ac{fl}-specific Leaf~\cite{caldas1812leaf} benchmark.
CIFAR10 and CIFAR100 both consist of~$|\mathcal{D}| = 50,000$ data samples which are distributed equally in quantity to the devices. A test dataset that consists of~$10,000$ samples is used for evaluating the accuracy. CIFAR10 consists of~$J=10$ classes, while CIFAR100 has~$J=100$ classes. Each input image has a resolution of~$3\times32\times32$ pixels. TinyImageNet uses a~$3\times64\times64$ resolution and consists of~$|\mathcal{D}| = 100,000$ samples with~$J=200$ classes. FEMNIST consists of~$|\mathcal{D}| = 640,500$ grayscale training images with~$J=62$ classes. For evaluation $16,000$ samples are used. For compatibility, all images are upscaled to~$3\times32\times32$.
Both CIFAR10 and CIFAR100 are evaluated on ResNet20~\cite{he2015deep} (adapted version for CIFAR datasets), whereas FEMNIST on ResNet8.
Data is distributed to each device~$c$, such that an equally-sized data subset~$\mathcal{D}_c$ is allocated per device.
In each round, a subset of devices~$\mathcal C^{(r)}$ is randomly selected for participation.
In the case of CIFAR10/100, we select~$|\mathcal{C}_c|=10$ with a total of~$|\mathcal{C}|=100$, in case of TinyImageNet, we use~$|\mathcal{C}|=200$, whereas for FEMNIST we select~$|\mathcal{C}|=3550$ (in line with Leaf).
We train with an initial learning rate of~$\eta=0.1$, batch size of~$32$, sinuous decay of~$\eta = 0.001$, and a total of~$R=1000$ rounds.
In each round, participating devices iterate once over their local dataset.
We report the average accuracy after~$R$ rounds of training, as well as the standard deviation of~$3$ independent random seeds. We apply data augmentation techniques for training such as random cropping, and horizontal and vertical flips (the last two are omitted for FEMNIST).
Finally, to account for accuracy degradation due to hardware approximation, we build on top of ApproxTrain~\cite{gong2022approxtrain} using custom CUDA GEMM kernels for approximation-aware convolution operations.
We evaluate our approach using a custom \ac{fl} framework based on TensorFlow~\cite{tensorflow2015}, as described in \cref{sec:technique}. 

\textbf{Device heterogeneity:} 
Our realistic \ac{fl} setup assumes that the devices' energy budgets for training are heterogeneous. 
Specifically, we separate devices into two groups: the first group, which makes~$\frac{1}{3}$ of all devices, uses the baseline accelerator C1, while~$\frac{2}{3}$ of the devices use an approximate accelerator (C2-C5). Hence, for our experiments with CIFAR10 and CIFAR100, devices~$[0,\ldots,29]$ use C1, while~$[30,\ldots,99]$ use C2-C5. 

\begin{figure}
    \centering
    \includegraphics[page=1]{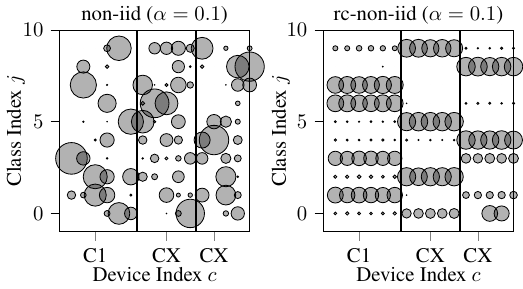} 
        \caption{Exemplary visualization of non-\ac{iid} (left) and rc-non-\ac{iid} (right) distributions for a total of 16 devices and 10 classes. Devices are equally grouped into three groups, where group one uses C1, and group 2 and 3 use CX (or SX in the case of the baselines) as described in our setup. The size of the circle represents the quantity of data samples of a specific class on a specific device. It can be seen on the right, contrary to the standard non-\ac{iid} on the left, that in this case there is a correlation between groups (i.e., device resources) and the quantity of class-specific data samples.}
    \label{fig:distribution}
\end{figure}

\textbf{Data heterogeneity:}
We distribute the classes~$J$ in~$\mathcal{D}$ in two ways. The first is in a typical non-\ac{iid} fashion. We generate this by using a Dirichlet distribution~\cite{hsu2019measuring}, where the degree of non-\ac{iid}-ness can be controlled with $\alpha$ (set to~$\alpha=0.1$). The resulting mapping of classes to devices is visualized in~\cref{fig:distribution} (left). The size of the circles represents the quantity of data for a specific class. It can be seen that the distribution of classes does not correlate with the device groups.

Secondly, to evaluate whether constrained devices can make a contribution to the global model, we distribute data in a resource-correlated non-\ac{iid} fashion, where we correlate data with the device energy budget.
Data is split into three groups, labeled~$\mathcal{D}_{g1}, \mathcal{D}_{g2}, \mathcal{D}_{g3}$, to have groups of equal sizes. 
Further, we map energy budgets and data using $[\{\text{C1},\mathcal{D}_{g1}\}, \{\text{CX},\mathcal{D}_{g2}\}, \{\text{CX},\mathcal{D}_{g3}\}]$ (X $\in [1,2,3,4,5]$). Devices within a group, sample data from~$\mathcal{D}_{g}$ in an \ac{iid} fashion. However, some class samples are only available in specific groups, requiring devices in that group to embed their class knowledge into the \ac{fl} model. This is visualized in \cref{fig:distribution} (right).
To evaluate if devices that use approximate accelerators can make a contribution to the global model, we evaluate the group-accuracy of devices in group~$g3$ (w.r.t.~$\mathcal{D}_{g3}$) according to \cref{eq:precision_recall} and \cref{eq:f1}.

\textbf{Fairness metric:}
To assess fairness, we examine the performance of the global model on the \emph{resource-group specific data distribution}~$\mathcal{D}_{g}$. This involves calculating the per-class accuracy for each class~$j$ using:

\begin{align}
    \label{eq:precision_recall}
    \text{accuracy}_j = \frac{\text{TP}_j}{\text{TP}_j + \text{TN}_j + \text{FP}_j + \text{FN}_j},
\end{align}

where~$\text{TP}_j$ denotes the number of \enquote{true positives} for class~$j$, and~$\text{TN}_j$,~$\text{FP}_j$, and~$\text{FN}_j$ represent the \enquote{true negatives,} \enquote{false positives,} and \enquote{false negatives} with respect to class~$j$, respectively.

We then compute the \emph{group-accuracy} by weighting the class-specific accuracy scores by the occurrence of samples in~$\mathcal{D}_{g}$ (i.e., where the label is~$y=j$), given by:

\begin{align}
    \label{eq:f1}
    \text{group-accuracy} = \frac{1}{|\mathcal{D}_g|} \sum_{j=0}^{J-1} |\mathcal{D}_{g, y=j}| \cdot \text{accuracy}_j,
\end{align}

where~$\mathcal{D}_g$ represents a combined dataset of devices within a specific group. We consider \ac{fl} training to be fair if the variance in the group-accuracy across the groups~$g1$, $g2$, and~$g3$ is low, indicating accuracy parity.

\subsection{State of the art comparison}
We compare our technique against two approaches:

\textbf{HeteroFL}: HeteroFL~\cite{diao2020heterofl} is a state-of-the-art algorithmic technique that lowers the resources in training. 
Specifically, a device incapable of training the full server \ac{nn} receives a downscaled model. 
Every layer in the \ac{nn} is downscaled using a scale factor~$s \in (0,1]$,
such that the downscaled parameters of a layer (i.e., a convolutional layer)
are~$w_{\text{downscaled}}
\in
\mathbb{R}^{
\lfloor s P  \rfloor \times \lfloor s Q \rfloor}$,
where~$P$ labels the input dimension and~$Q$ the output dimension.
Thereby, when $s$ decreases, the number of \ac{mac} operations is lowered quadratically by~$\sim s^2 PQ$.
We evaluate the energy consumption similarly to our technique for~$s \in [1.0, 0.5, 0.25, 0.125]$, which we label S1, S2, S3, and S4.
Even though analytically the number of \acp{mac} lower quadratically, we observe that using an accelerator, the small subsets of filters can not be efficiently tiled anymore, causing the systolic array to pad dimensions with zeros, effectively wasting energy. Therefore, we observe that compared to a CPU, where operations are performed sequentially, the assumption of quadratic reduction does not hold true in the case of GPUs or systolic accelerators. For a fair comparison, we evaluate HeteroFL on the same full-precision accelerator we use for our hardware-level approach, as HeteroFL is hardware agnostic and a CPU-based comparison would result in significantly higher energy usage~\cite{9730377}. The energy breakdown of the aforementioned scaling configurations is presented in \cref{fig:energy_components}(right). We observe that an analytic reduction of the \ac{nn} parameters by~$8\times$ results only in a~$2.34\times$ energy reduction.

\textbf{FedRolex}: Similar to HeteroFL, FedRolex~\cite{alam2022fedrolex} utilizes subsets, i.e. the resource reduction mechanism is the same as with HeteroFL. However, different from HeteroFL, constrained devices do not statically train the same part of the full server \ac{nn} throughout the training. Instead, devices that train a subset, iterate over the full parameters in a rolling window fashion. Specifically, a downscaled layer (i.e., a convolutional layer) $w_{\text{downscaled}}
\in
\mathbb{R}^{
\lfloor s P  \rfloor \times \lfloor s Q \rfloor}$ is created every round $r$ by using $\hat r = r \ \text{mod}\ Q$ to create the reduced layer's output indices~$I$ using
\begin{align}
\label{eq:fedrolex}
    I = \begin{cases} 
    \{ \hat r, \hat r+1, \ldots, r + \lfloor s Q \rfloor -1 \} & \text{if } \hat r + \lfloor s Q \rfloor \leq Q \\
    \{ \hat r, \hat r+1, \ldots, Q -1 \} \ \cup \\
    \{0,\ldots, \hat r + \lfloor s Q  \rfloor - 1 - Q \} & \text{otherwise}
    \end{cases}.
\end{align}
The layer's input indices are determined by the preceding layer's output indices.

\textbf{Small model}: This is a naive baseline that enforces homogeneous \ac{fl} training by using a small model of low enough energy requirements, such that all devices are capable of training the model. 
We emulate the baseline by scaling down the \ac{nn} similarly to HeteroFL. However, differently to HeteroFL, all participating devices train the small model.

For the evaluation of baselines, we map energy budgets and data using $[\{\text{S1},\mathcal{D}_{g1}\}, \{\text{SX},\mathcal{D}_{g2}\}, \{\text{SX},\mathcal{D}_{g3}\}]$ (X $\in [1,2,3,4]$), where S1 refers to the full \ac{nn} model, which is equivalent to C1.

\textbf{Drop Devices:} As a lower bound for assessing fairness, we analyze the effects of excluding constrained devices from the training process. Specifically, devices in groups $\{\text{SX}, \mathcal{D}_{g2}\}$ and $\{\text{SX}, \mathcal{D}_{g3}\}$ are excluded from training, reflecting the current production deployment scenario~\cite{yang2018applied}.

\textbf{FedProx:} Lastly, we compare our technique against FedProx~\cite{li2020federated}, a widely adopted method to tackle heterogeneity in \ac{fl}. FedProx shares some similarities with dropping devices; however, in FedProx, constrained devices are not dropped but instead train fewer mini-batches per round. Additionally, FedProx applies a norm that keeps the local weights close to the global weights during training. While the assessment of energy reduction is straightforward (i.e., training with 50\% of the mini-batches results in 50\% of the energy used), the norm used in FedProx requires additional backpropagation steps and maintaining the server weights in memory. This cannot currently be mapped to our accelerator. Hence, for a fair comparison, we assume it to have zero cost. We study training with $100\%$, $50\%$, and $25\%$ of mini-batches, which we label F1 ($1\times$), F2 ($2\times$), and F3 ($4\times$). For all experiments, $\mu=0.01$ is used, as it produces the best results across all evaluated datasets.
\section{Experimental Evaluation}
\label{sec:experimental_evaluation}

\begin{figure*}
    \centering
    \includegraphics[page=1]{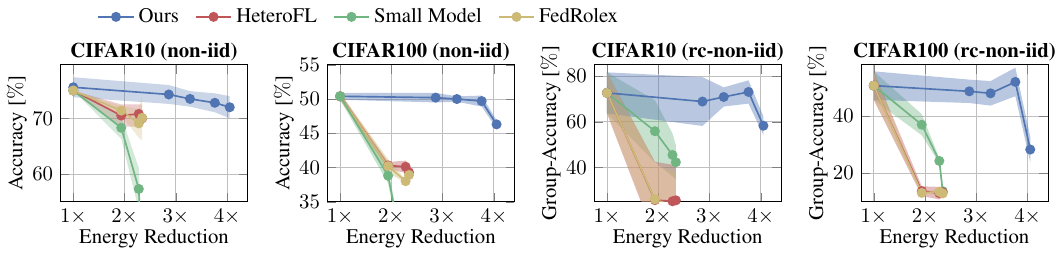} 
\caption{Average accuracy and group-accuracy evaluation for CIFAR10/100 and ResNet20 with 3 independent seeds for all studied techniques. Our design-time technique solutions populate a significantly higher Pareto front of accuracy/energy trade-offs compared to the state of the art.
}
\label{fig:accuracy}
\end{figure*}

We evaluate our technique in terms of energy reduction and training classification accuracy over the aforementioned datasets and energy-constrained scenarios (i.e., C1-C5).
\cref{fig:accuracy} visualizes the accuracy/energy trade-off compared to the state-of-the-art, where the mean and standard deviation of the accuracy of~$3$ independent runs for each configuration are given.
Overall, solutions from our design-time technique populate a significantly higher Pareto front of accuracy/energy trade-offs compared to the state of the art.
Additionally for completeness, we analytically report our findings in \cref{tab:accuracy}.
An analysis of the two types of class distribution follows.
\begin{table*}[ht]
    \renewcommand{\arraystretch}{1.125}

    \centering
    \caption{
    Top-1 classification accuracy in \% (group-accuracy for rc-non-\ac{iid}) of our technique in comparison with the state of the art.
    Both heterogeneity scenarios (i.e., non-iid and rc-non-iid) are evaluated with each dataset.
    Our technique employs the accelerator configurations C1-C5, whereas the state of the art scales the \ac{nn} model instead (S1-S4).
    The energy scaling achieved by each configuration is enclosed in parentheses.
    Overall, our technique allows for up to 4$\times$ energy reduction with minimal accuracy degradation. In total, we run 348 experiments which accumulates to $\sim72$ days of simulation on NVIDIA V100 GPUs.
    }
\begin{tabular}{l l l  | c c c c c c c c}

    \toprule
    \multicolumn{2}{l}{Setting} & Technique & C1/S1 (1$\times$) & S2 (1.94$\times$) & S3 (2.28$\times$) & S4 (2.34$\times$) & C2 (2.86$\times$)  & C3 (3.28$\times$) & C4 (3.76$\times$) & C5 (4.05$\times$)\\
    \midrule
    \multirow{8}{*}{\rotatebox{90}{non-\ac{iid}}}
     & \multirow{4}{*}{\rotatebox{90}{\small{CIFAR10}}} & \textbf{Ours} &
     $|$  & 
    - &
    - &
    - &
    74.3 $\pm$ 1.7 &
    73.5 $\pm$ 1.4 &
    72.8 $\pm$ 1.7 &
    72.0 $\pm$ 2.0 \\

    &  &HeteroFL~\cite{diao2020heterofl} &
    75.6 $\pm$ 1.8 &
    70.0 $\pm$ 2.0 &
    69.2 $\pm$ 2.4 &
    70.1 $\pm$ 1.6 &
    - &
    - &
    - &
    - \\

    &  &FedRolex~\cite{alam2022fedrolex} &
    $|$ &
    71.4 $\pm$ 0.8 &
    69.1 $\pm$ 2.6 &
    70.1 $\pm$ 4.1 &
    - &
    - &
    - &
    - \\

    &  &Small Model &  $|$ &
    68.3 $\pm$ 2.0 &
    57.3 $\pm$ 3.8 &
    43.4 $\pm$ 1.8 &
    - &
    - &
    - &
    - \\
    \cmidrule{2-11}

    &  \multirow{4}{*}{\rotatebox{90}{\small{CIFAR100}}}  & \textbf{Ours} &
    $|$ &
    - &
    - &
    - &
    50.2 $\pm$ 0.7 &
    50.0 $\pm$ 0.4 &
    49.7 $\pm$ 0.8 &
    46.3 $\pm$ 0.6 \\

    &  & HeteroFL~\cite{diao2020heterofl} &
    50.4 $\pm$ 0.5 &
    40.3 $\pm$ 0.5 &
    40.1 $\pm$ 0.9 &
    39.3 $\pm$ 1.2 &
    - &
    - &
    - &
    - \\
    
    &  &FedRolex~\cite{alam2022fedrolex} &
    $|$ &
    40.2 $\pm$ 1.0 &
    38.0 $\pm$ 0.2 &
    38.9 $\pm$ 1.1 &
    - &
    - &
    - &
    - \\

    &  & Small Model &
    $|$ &
    38.8 $\pm$ 1.1 &
    25.7 $\pm$ 0.8 &
    13.7 $\pm$ 1.7 &
    - &
    - &
    - &
    - \\

    \midrule

    \multirow{10}{*}{\rotatebox{90}{rc-non-\ac{iid}}}
 & \multirow{5}{*}{\rotatebox{90}{\small{CIFAR10}}}  & \textbf{Ours} &
    $|$ &
    - &
    - &
    - &
    68.9 $\pm$ 10.7 &
    70.9 $\pm$ \phantom{0}4.2 &
    73.1 $\pm$ \phantom{0}5.1 &
    58.3 $\pm$ \phantom{0}4.1 \\

    &  & HeteroFL~\cite{diao2020heterofl} &
    72.7 $\pm$ 9.0 &
    25.7 $\pm$ 16.8 &
    25.2 $\pm$ 16.0 &
    25.6 $\pm$ 16.0 &
    - &
    - &
    - &
    - \\

    &  &FedRolex~\cite{alam2022fedrolex} &
    $|$ &
    26.0 $\pm$ 15.7 &
    23.3 $\pm$ 14.5 &
    23.3 $\pm$ 15.4 &
    - &
    - &
    - &
    - \\

    &  & Small Model &
    $|$ &
    55.9 $\pm$ 13.6 &
    45.5 $\pm$ \phantom{0}9.4 &
    42.3 $\pm$ \phantom{0}7.5 &
    - &
    - &
    - &
    - \\

    & & Drop Devices &
        $|$ &
    23.0 $\pm$ 15.1 &
    23.0 $\pm$ 15.1 &
    23.0 $\pm$ 15.1 &
    23.0 $\pm$ 15.1 &
    23.0 $\pm$ 15.1 &
    23.0 $\pm$ 15.1 &
    23.0 $\pm$ 15.1\\

    \cmidrule{2-11}

    &  \multirow{5}{*}{\rotatebox{90}{\small{CIFAR100}}} & \textbf{Ours} &
    $|$ &
    - &
    - &
    - &
    48.7 $\pm$ 4.1 &
    48.0 $\pm$ 4.3 &
    52.0 $\pm$ 5.0 &
    28.3 $\pm$ 4.5 \\

    &  & HeteroFL~\cite{diao2020heterofl} &
    50.7 $\pm$ 5.0 &
    13.7 $\pm$ 1.7 &
    13.0 $\pm$ 2.2 &
    13.3 $\pm$ 1.7 &
    - &
    - &
    - &
    - \\

    &  &FedRolex~\cite{alam2022fedrolex} &
    $|$ &
    13.1 $\pm$ 0.2 &
    13.5 $\pm$ 1.1 &
    12.9 $\pm$ 0.8 &
    - &
    - &
    - &
    - \\

    & & Small Model &
    $|$ &
    37.0 $\pm$ 3.7 &
    24.3 $\pm$ 0.9 &
    13.7 $\pm$ 3.1 &
    - &
    - &
    - &
    - \\

    & & Drop Devices &
        $|$ &
    13.2 $\pm$ 0.6 &
    13.2 $\pm$ 0.6 &
    13.2 $\pm$ 0.6 &
    13.2 $\pm$ 0.6 &
    13.2 $\pm$ 0.6 &
    13.2 $\pm$ 0.6 &
    13.2 $\pm$ 0.6 \\

    \cmidrule{1-11}
 \multicolumn{2}{l}{Setting} & Technique & C1/S1 (1$\times$) & S2 (1.96$\times$) & S3 (2.3$\times$) &S4 (2.37$\times$) & C2 (2.88$\times$)  & C3 (3.29$\times$) & C4 (3.77$\times$) & C5 (4.06$\times$)\\
     \cmidrule{1-11}
     \multirow{4}{*}{\rotatebox{90}{non-\ac{iid}}}
   &  \multirow{4}{*}{\rotatebox{90}{\small{FEMNIST}}}  & \textbf{Ours} &
    $|$ &
    - &
    - &
    - &
    82.2 $\pm$ 0.7 &
    82.1 $\pm$ 0.7 &
    81.4 $\pm$ 0.6 &
    78.7 $\pm$ 0.2 \\

    &  & HeteroFL~\cite{diao2020heterofl} &
    83.7 $\pm$ 0.5 &
    81.1 $\pm$ 0.3 &
    80.1 $\pm$ 0.8 &
    80.0 $\pm$ 0.9 &
    - &
    - &
    - &
    - \\

    &  &FedRolex~\cite{alam2022fedrolex} &
    $|$ &
   77.9 $\pm$ 0.3 &
    78.3 $\pm$ 1.3 &
    79.1 $\pm$ 1.7 &
    - &
    - &
    - &
    - \\

    &  & Small Model &
    $|$ &
    80.3 $\pm$ 0.5 &
    72.7 $\pm$ 0.4 &
    56.9 $\pm$ 4.0 &
    - &
    - &
    - &
    - \\
    \cmidrule{1-11}
    \multirow{5}{*}{\rotatebox{90}{rc-non-\ac{iid}}}
    &  \multirow{5}{*}{\rotatebox{90}{\small{FEMNIST}}} & \textbf{Ours} &
    $|$ &
    - &
    - &
    - &
    64.5 $\pm$ 7.7 &
    41.7 $\pm$ 6.1 &
    43.1 $\pm$ 8.4 &
    35.9 $\pm$ 7.4 \\

    &  & HeteroFL~\cite{diao2020heterofl} &
    80.2 $\pm$ 4.5 &
    38.9 $\pm$ 11.0 &
    37.6 $\pm$ 10.8 &
    37.4 $\pm$ 10.1 &
    - &
    - &
    - &
    - \\

    &  &FedRolex~\cite{alam2022fedrolex} &
    $|$ &
    40.7 $\pm$ 10.1 &
    35.7 $\pm$ \phantom{0}9.5 &
    35.2 $\pm$ \phantom{0}9.5 &
    - &
    - &
    - &
    - \\

    & & Small Model &
    $|$ & 
    80.0 $\pm$ \phantom{0}2.8 &
    71.5 $\pm$ \phantom{0}5.7 &
    48.2 $\pm$ \phantom{0}9.1 &
    - &
    - &
    - &
    - \\

    & & Drop Devices &
        $|$ &
    32.4 $\pm$ \phantom{0}8.9 &
    32.4 $\pm$ \phantom{0}8.9 &
    32.4 $\pm$ \phantom{0}8.9 &
    32.4 $\pm$ 8.9 &
    32.4 $\pm$ 8.9 &
    32.4 $\pm$ 8.9 &
    32.4 $\pm$ 8.9 \\
    \cmidrule{1-11}
 \multicolumn{2}{l}{Setting} & Technique & C1/S1 (1$\times$)& S2 (2.02$\times$) & S3 (2.42$\times$) & S4 (2.50$\times$) & C2 (2.92$\times$)  & C3 (3.36$\times$) & C4 (3.88$\times$) & C5 (4.20$\times$)\\
     \cmidrule{1-11}
\multirow{4}{*}{\rotatebox{90}{non-\ac{iid}}}
& \multirow{4}{*}{\rotatebox{90}{\small{TinyINet}}}  & \textbf{Ours}&
  $|$ & - & - & - & 28.0 $\pm$0.2 & 27.1$\pm$0.7 & 28.2$\pm$0.2 & 25.5$\pm$0.2\\

& & HeteroFL~\cite{diao2020heterofl} & 30.5 $\pm$ 1.1 & 25.5 $\pm$ 0.2 & 23.4 $\pm$ 0.1 & 23.9 $\pm$ 0.4 & - & - & - & -\\
& & FedRolex~\cite{alam2022fedrolex} & $|$ & 23.3 $\pm$ 0.8 & 22.5 $\pm$ 0.5 & 23.5 $\pm$ 0.7 &- &- &- &-\\
& & Small Model & $|$ &23.5 $\pm$ 0.1 &17.4 $\pm$ 1.2 & 11.1 $\pm$ 0.6 & - & - & -\\

\cmidrule{1-11}    
     \multirow{5}{*}{\rotatebox{90}{rc-non-\ac{iid}}}
& \multirow{5}{*}{\rotatebox{90}{\small{TinyINet}}}  & \textbf{Ours} &
    $|$&  - & - & - & 27.6 $\pm$ 9.1 & 24.0 $\pm$ 7.7 & 23.5 $\pm$ 7.5 & 8.4 $\pm$ 2.4 \\

& & HeteroFL~\cite{diao2020heterofl} & 26.9 $\pm$ 6.5 & \phantom{0}6.2 $\pm$ 0.9 & \phantom{0}7.1 $\pm$ 1.4 &  \phantom{0}6.8 $\pm$ 0.3 & - & - & - & -\\
& & FedRolex~\cite{alam2022fedrolex}& $|$ & \phantom{0}5.8 $\pm$ 0.3 & \phantom{0}6.2 $\pm$ 0.7 & \phantom{0}6.4 $\pm$ 1.3 & - & - & - & -\\
& & Small Model & $|$ & 20.3 $\pm$ 7.2 & 15.9 $\pm$ 5.8 & \phantom{0}8.9 $\pm$ 5.6 & - & - & - & -\\
& & Drop Devices & $|$ & \phantom{0}7.4 $\pm$ 0.8 &  \phantom{0}7.4 $\pm$ 0.8 & \phantom{0}7.4 $\pm$ 0.8 & \phantom{0}7.4 $\pm$ 0.8 &  \phantom{0}7.4 $\pm$ 0.8 &  \phantom{0}7.4 $\pm$ 0.8 &  \phantom{0}7.4 $\pm$ 0.8\\
\bottomrule
\end{tabular}
\label{tab:accuracy}
\end{table*}

\textbf{Mainline non-\ac{iid} results:}
For non-\ac{iid} distribution and CIFAR10, we observe that using approximate accelerators has only a minor effect on the global accuracy while being able to reduce the resources of the constrained devices drastically.
Specifically, when we deploy the approximate accelerator C3 instead of C1, devices require~$\sim 2.8\times$ less energy for training but the overall \ac{fl} accuracy is only lowered by~$\SI{2.1}{\pp}$.
If we compare C1 to C5, we observe a drop of~$\SI{3.6}{\pp}$, but devices are able to lower their training energy by~$\sim 4.0\times$. 
Compared to our technique, we observe that HeteroFL only allows for limited energy reduction ($\sim 2.3\times$ with S4 compared to the full model S1).
Using S2, S3, and S4 the accuracy drops accordingly by~$\SI{4.5}{\pp}$. Similar behavior can be observed for FedRolex.
It can be observed that when using a smaller model than S1 (small model baseline), the remaining filters do not provide a sufficient amount of model capacity to reliably predict the respective classes in the training set, hence the accuracy drops by up to~$\SI{31,6}{\pp}$ for S4.

In general, the trends for CIFAR100 are similar to CIFAR10.
However, we notice that the drop of HeteroFL and FedRolex is larger (i.e.,~$\SI{10.1}{\pp}$ with S2 using HeteroFL) compared to our technique. Only by using C5, the accuracy drops more significantly by~$\SI{3.9}{\pp}$.
As CIFAR100 classification is a more challenging task than CIFAR10, we observe that the accuracy due to the limited capacity drops even further when using S2.

\textit{In essence, for non-\ac{iid} we observe that by using approximate accelerators in \ac{fl}, energy-constrained devices can better retain the accuracy and utilize a higher energy reduction compared to using heterogeneous subsets, as it is done in HeteroFL or FedRolex, and also compared to a homogeneous setting with a small model.}

\begin{figure}
    \centering
    \includegraphics[page=1]{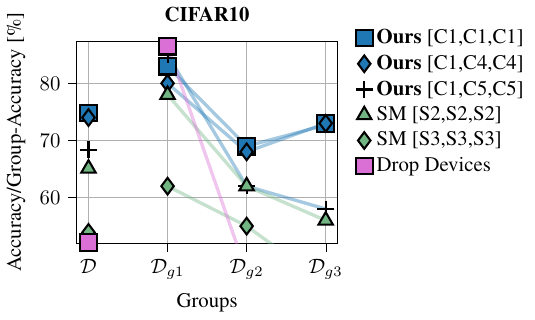} 
    \caption{Group-accuracy for group $g_1$ ($\mathcal{D}_{g1}$), $g_2$ ($\mathcal{D}_{g2}$), and $g_3$ ($\mathcal{D}_{g3}$), as well as average accuracy (evaluated on $\mathcal D$) for rc-non-\ac{iid} training with CIFAR10. The individual group-accuracies suggest that even when no approximation is applied, the individual accuracy of the groups is not equal as the the data subsets can vary in detection complexity. 
    Further, it can be observed that using approximate accelerators can retain the group-accuracy completely (i.e., group~$g_2$ and~$g_3$ using~C4), however, we observe that comes with a lower group-accuracy of~C1, hence, the average accuracy when using~C4 is lower. Additionally, we observe, that with a strong approximation degree (i.e.,~C5), the influence of the approximate accelerator is reduced, hence group-accuracy of group~$g_2$ and~$g_3$ is reduced, while the group-accuracy of the full precision group is higher. However, in this case, the average accuracy is lower. SM refers to the small model baseline.}
    \label{fig:rcnoniid}
\end{figure}

\textbf{Mainline rc-non-\ac{iid} results:}
For CIFAR10, we observe that using approximate accelerators, energy-constrained devices can make a meaningful contribution to the global model, as the group-accuracy is only minimally impacted by approximation.
Additionally, the large gap in \cref{tab:accuracy} between the complete dropping of devices (Drop Devices) and our technique highlights this.
Specifically, using C2 only lowers the group-accuracy by~$\SI{3.8}{\pp}$. Counterintuitively, it can be observed that using C3 improves the group-accuracy, even compared to C1. We think that is caused by noise and quantization effects of the constrained devices, degrading the contribution of the non-constrained devices as constrained devices make up for 66\% of all devices. However, as non-\ac{iid} results show, overall global accuracy is still degraded by C3.
For C5 (i.e., the most aggressive approximation), it is evident that the approximation has a greater impact on group-accuracy compared to C1, resulting in a decrease of $\SI{14.4}{\pp}$ in accuracy. This observation is illustrated in \cref{fig:rcnoniid}, which depicts the group-accuracies for all three groups as well as the average accuracy ($\mathcal{D}$). The figure demonstrates that higher levels of approximation can increase group-accuracy, albeit at the expense of average accuracy (i.e., the accuracy of the non-constrained group decreases). Generally, it is observed that approximation achieves high average accuracy and lower accuracy variance among the groups, promoting accuracy parity (fairness) among them. In contrast, dropping devices causes the global model to be strongly biased towards data from group $g_1$ and fails to represent groups $g_2$ and $g_3$. The small model baseline exhibits relatively low variance but also lower overall accuracy due to its limited model capacity.

For HeteroFL however, we observe that constrained devices completely fail to have their specific classes represented in the global \ac{fl} model, since the accuracy drops by~$\SI{37}{\pp}$ when using S2 and remains to such low levels for S3 and S4. Contrary to the non-\ac{iid} case, a common homogeneous model better retains the group-accuracy, however, the capacity problem of the small model in general persists.

We observe the same trends for CIFAR100.
It can be seen, that group-accuracy with C5 is lower compared to CIFAR10. The drop in accuracy is thereby comparable to a small model with S3.
However, it has to be noted that using the approximate configuration C5 reduces the training energy by~$\sim 4\times$, while a small model with S3 only reduces the training energy by~$\sim 2.2\times$.

\textit{In summary, we observe that using approximate accelerators for energy-constrained devices allows devices to make a meaningful contribution to the global model despite having less energy available for training, which is essential to maintain fairness in the global model. It can be observed that in HeteroFL and FedRolex, devices that train a subset hardly incorporate their classes in the global model, effectively having no impact on the model. The usage of small homogeneous models maintains fairness slightly better but has a lower accuracy due to the limited model capacity.}

\textbf{Robustness to larger scale of devices}:
To evaluate our technique with other \ac{nn} configurations and a larger scale of devices, we evaluate ResNet8 with FEMNIST. We observe, that the total energy for C1 is reduced to~$\SI{0.10}{\joule}$ per mini-batch compared to ResNet20 ($\SI{0.21}{\joule}$). However, the energy scaling of C2-C5 and S2-S4 remains in a very similar range. In non-\ac{iid} training, using approximate accelerators, we reach accuracies of~$83.7\%\pm 0.5\%$ for C1/S1, with a minimal drop in accuracy using approximate accelerators in the case of non-iid. Similarly, to CIFAR10, HeteroFL and FedRolex lose more in accuracy compared to our technique. Counterintuitively, we observe for FedRolex that when smaller subsets are used e.g., S4, the accuracy is better retained. A potential reason for that is that larger subsets of groups~$g_2$ and~$g_3$ interfere more with the full model training of group~$g_1$.

For rc-non-\ac{iid}, we observe that our technique retrains higher group-accuracy compared to the state of the art, however, in general all techniques relatively lose more accuracy compared to CIFAR10 and CIFAR100.

\textit{The experiments with FEMNIST and ResNet8 show similar trends to CIFAR10/100, further demonstrating the robustness of our technique against different datasets and \ac{nn} configurations.}

\textbf{Comparison with FedProx:} We compare our technique with FedProx, which achieves 74.1\% $\pm$ 2.1\% and 73.0\% $\pm$ 2.0\% for F2 and F3 on non-\ac{iid} CIFAR10, and 46.7\% $\pm$ 0.6\% and 42.6\% $\pm$ 0.9\% for non-\ac{iid} CIFAR100. While the results for non-\ac{iid} CIFAR10 are on par with our technique, non-\ac{iid} CIFAR100 shows significantly worse accuracy compared to our method. We assume that this is caused by the lower per-class sample count. A similar trend can be observed with rc-non-\ac{iid} CIFAR10 and CIFAR100 (group-accuracy), where FedProx achieves 64.6\% $\pm$ 5.2\% (F2), 48.6\% $\pm$ 9.5\% (F3), and 38.4\% $\pm$ 6.5\% (F2) and 21.3\% $\pm$ 0.9\% (F3), respectively. Especially in rc-non-\ac{iid} environments, FedProx with F3 ($4\times$) deteriorates significantly compared to our technique. From this, we conclude that it is better to train with more mini-batches using our approximate arithmetic than to train with fewer mini-batches at high precision.

\textbf{Robustness to larger scale datasets:}
We evaluate TinyImageNet to ensure that our technique scales to larger, more complex datasets. TinyImageNet contains twice as many data samples and twice as many classes as CIFAR100. Additionally, TinyImageNet increases the resolution to~$3\times64\times64$, a~$4\times$ increase compared to CIFAR. We observe in our energy model a similar increase in required energy, i.e., C1/S1 requires~\SI{0.88}{\joule} per mini-batch. C2-C5, as well as S2-S4 scale with~$2.9\times$, $3.4\times$, $3.9\times$, $4.2\times$, and~$2.0\times$, $2.4\times$, $2.5\times$, respectively. Generally, for non-\ac{iid} and rc-non-\ac{iid} scenarios, we observe behavior similar to that seen with the CIFAR datasets (\cref{tab:accuracy}). For C4, we observe higher accuracy compared to C3, which we attribute to the noise providing a regularization effect on the training, consistent with previous observations. Only with C5 for rc-non-\ac{iid}, do we see a stronger degradation in accuracy.

\begin{table*}[ht]
    \renewcommand{\arraystretch}{1.125}

    \centering
    \caption{Ablation study to test the robustness against a larger degree of heterogeneity (higher mixture of C1-C5). The results suggest that overall our technique is robust against such scenarios, as accuracy (average accuracy of 3 independent runs in \%) is not collapsing.}
\begin{tabular}{l | c c c c c}
\toprule
Setting & C1/S1 & \emph{Mix1} & \emph{Mix2} & \emph{Mix3} & \emph{Mix4}\\
Mixture & $\{\frac{5}{5}\text{C1}\}$ & $\{\frac{1}{5}\text{C1},\frac{4}{5}\text{C2}\}$ & $\{\frac{1}{5}\text{C1},\frac{1}{5}\text{C2}, \frac{3}{5}\text{C3}\}$& $\{\frac{1}{5}\text{C1},\frac{1}{5}\text{C2}, \frac{1}{5}\text{C3}, \frac{2}{5}\text{C4}\}$& $\{\frac{1}{5}\text{C1},\frac{1}{5}\text{C2}, \frac{1}{5}\text{C3}, \frac{1}{5}\text{C4}, \frac{1}{5}\text{C5}\}$\\
\midrule
non-iid CIFAR10 & 75.6 $\pm$ 1.8 & 74.6 $\pm$ 2.5 & 75.0 $\pm$ 1.1 & 74.6 $\pm$ 3.0 & 72.9 $\pm$ 2.2 \\
non-iid CIFAR100 & 50.4 $\pm$ 0.5 & 50.5 $\pm$ 1.6 &  51.2 $\pm$ 1.1 & 50.5 $\pm$ 0.9 & 49.9 $\pm$ 1.2\\
non-iid FEMNIST & 83.7 $\pm$ 0.5 & 80.3 $\pm$ 1.1 & 82.7 $\pm$ 0.5 & 81.4 $\pm$ 0.3 & 80.4 $\pm$ 0.8\\
\bottomrule
\end{tabular}
\label{tab:accuracy_mix}
\end{table*}
\textbf{Robustness to mixtures of approximation levels:} To verify that our technique is robust against a high degree of heterogeneity, we evaluate how the global model accuracy degrades with increasing heterogeneity. In the main experiments (\cref{tab:accuracy}), a mix of~$\frac{1}{3}\text{ C1}$ and~$\frac{2}{3}\text{ CX}$ was used. We present varying mixtures for CIFAR10, CIFAR100, and FEMNIST in \cref{tab:accuracy_mix}. Specifically, we start with a mix of~$\frac{1}{5}\text{ C1}$ and gradually increase the number of different approximation levels in the mixture, so that in \textit{Mix4}, each approximation level (C1-C5) is equally present in the training. For all three datasets, we observe that the global model accuracy does not collapse, and in some cases, a higher mixture of approximation levels can even slightly improve accuracy. We attribute this improvement to the noise, which provides a regularization effect.

\textbf{Robustness to mixture of techniques:}
To demonstrate that our technique is orthogonal to algorithm-level techniques to lower energy, we combine FedProx with our approximate accelerator. Specifically, we use only 50\% of the data per round (F2), as it only results in a moderate accuracy loss, and use a moderate approximation (C4). Combining both techniques achieves an energy reduction of $7.5\times$ (energy consumption of norm is assumed zero). For rc-non-\ac{iid} CIFAR10 and CIFAR100, we reach 67.7\% $\pm$ 2.7\% and 37.1\% $\pm$ 6.3\% accuracy, respectively. We conclude from that, that moderate dropping of data per round and moderate approximation outperforms more aggressive approximation levels (C5). This demonstrates that our hardware-based approach can be applied orthogonally to algorithm-level techniques.

\textbf{Robustness to different approximation techniques:}
To demonstrate that our accelerator design is not limited to \ac{mbm}~\cite{saadat2018minimally}, we replace the approximate mantissa multiplication with a Mitchell approximate multiplier (MIT)~\cite{mitchell1962computer}. We maintain the same truncation of weights and inputs as in \ac{mbm} for a fair comparison. Our synthesis of MIT-7, MIT-3, and MIT-1 results in slightly higher reductions in energy compared to their \ac{mbm} counterparts, achieving reductions of $3.4\times$, $4.3\times$, and $4.8\times$ respectively. With non-\ac{iid} CIFAR100, we achieve similar accuracy results, reaching $50.5\%\pm1.1\%$, $50.6\%\pm1.2\%$, and $48.7\%\pm1.0\%$ for MIT-7, MIT-3, and MIT-1, respectively.

\emph{In summary, our proposed technique is robust against a high degree of heterogeneity and robust against different mantissa approximation techniques.}
\section{Conclusion}
\label{sec:conclusion}
In this work, we are the first to tackle the device heterogeneity problem in \ac{fl} at design time.
We design and employ training-capable \ac{nn} accelerators to meet the diverse energy budgets of \ac{fl} devices.
Our designs are enhanced with compressed arithmetic formats and approximate computing, whilst our energy gains are accurately quantified with a compute- and memory-aware energy model.
Although we have not explicitly studied privacy guarantees in this work, we assume the added quantization noise does not weaken privacy and assume it to be on par with full precision federated averaging, as recent works have shown that quantization in transmission can have beneficial effects on privacy~\cite{10000632,kang2024effect}. Further, privacy can be enhanced by employing differential privacy techniques~\cite{wei2020federated}, which are orthogonal to our approach.
In contrast to the modest energy reduction achieved by the hardware-unaware state-of-the-art (i.e. only $2.3\times$), our technique yields up to~$4\times$ energy reductions, keeping accuracy considerably higher, whilst maintaining fairness for constrained devices.

\bibliographystyle{ieeetr}
\bibliography{bib/bibfile}

\begin{thebibliography}{10}

\bibitem{pfeiffer2023survey}
K.~Pfeiffer, M.~Rapp, R.~Khalili, and J.~Henkel, ``Federated learning for
  computationally constrained heterogeneous devices: A survey,'' {\em ACM
  Computing Surveys}, vol.~55, no.~14s, 2023.

\bibitem{yang2018applied}
T.~Yang, G.~Andrew, H.~Eichner, H.~Sun, W.~Li, N.~Kong, D.~Ramage, and
  F.~Beaufays, ``Applied federated learning: Improving google keyboard query
  suggestions,'' {\em arXiv:1812.02903}, 2018.

\bibitem{maeng2022towards}
K.~Maeng, H.~Lu, L.~Melis, J.~Nguyen, M.~Rabbat, and C.-J. Wu, ``Towards fair
  federated recommendation learning: Characterizing the inter-dependence of
  system and data heterogeneity,'' in {\em Proceedings of the 16th ACM
  Conference on Recommender Systems}, 2022.

\bibitem{diao2020heterofl}
E.~Diao, J.~Ding, and V.~Tarokh, ``Heterofl: Computation and communication
  efficient federated learning for heterogeneous clients,'' in {\em
  International Conference on Learning Representations (ICLR)}, 2020.

\bibitem{horvath2021fjord}
S.~Horvath, S.~Laskaridis, M.~Almeida, I.~Leontiadis, S.~Venieris, and N.~Lane,
  ``Fjord: Fair and accurate federated learning under heterogeneous targets
  with ordered dropout,'' {\em Advances in Neural Information Processing
  Systems}, vol.~34, pp.~12876--12889, 2021.

\bibitem{alam2022fedrolex}
S.~Alam, L.~Liu, M.~Yan, and M.~Zhang, ``Fedrolex: Model-heterogeneous
  federated learning with rolling sub-model extraction,'' in {\em Advances in
  Neural Information Processing Systems}, 2022.

\bibitem{caldas2018expanding}
S.~Caldas, J.~Kone{\v{c}}ny, H.~B. McMahan, and A.~Talwalkar, ``Expanding the
  reach of federated learning by reducing client resource requirements,'' {\em
  arXiv:1812.07210}, 2018.

\bibitem{cheng2022does}
G.~Cheng, Z.~Charles, Z.~Garrett, and K.~Rush, ``Does federated dropout
  actually work?,'' in {\em Proceedings of the IEEE/CVF Conference on Computer
  Vision and Pattern Recognition}, pp.~3387--3395, 2022.

\bibitem{pfeiffer2023cocofl}
K.~Pfeiffer, M.~Rapp, R.~Khalili, and J.~Henkel, ``Coco{FL}: Communication- and
  computation-aware federated learning via partial {NN} freezing and
  quantization,'' {\em Transactions on Machine Learning Research}, 2023.

\bibitem{Chen:SSC:2017:eyeriss}
Y.-H. Chen, T.~Krishna, J.~S. Emer, and V.~Sze, ``Eyeriss: An energy-efficient
  reconfigurable accelerator for deep convolutional neural networks,'' {\em
  IEEE Journal of Solid-State Circuits}, vol.~52, no.~1, 2017.

\bibitem{Armen:ACMSurv:2023:ac}
G.~Armeniakos, G.~Zervakis, D.~Soudris, and J.~Henkel, ``Hardware approximate
  techniques for deep neural network accelerators: A survey,'' {\em ACM
  Computing Surve}, vol.~55, no.~4, 2022.

\bibitem{baghersalimi2021personalized:r2_rebuttal}
S.~Baghersalimi, T.~Teijeiro, D.~Atienza, and A.~Aminifar, ``Personalized
  real-time federated learning for epileptic seizure detection,'' {\em IEEE
  journal of biomedical and health informatics}, vol.~26, no.~2, pp.~898--909,
  2021.

\bibitem{baghersalimi2023decentralized:r3_rebuttal}
S.~Baghersalimi, T.~Teijeiro, A.~Aminifar, and D.~Atienza, ``Decentralized
  federated learning for epileptic seizures detection in low-power wearable
  systems,'' {\em IEEE Transactions on Mobile Computing}, 2023.

\bibitem{tpu_v1}
N.~P. Jouppi, C.~Young, N.~Patil, D.~Patterson, G.~Agrawal, R.~Bajwa, S.~Bates,
  S.~Bhatia, N.~Boden, A.~Borchers, {\em et~al.}, ``In-datacenter performance
  analysis of a tensor processing unit,'' in {\em Proceedings of the 44th
  annual international symposium on computer architecture}, pp.~1--12, 2017.

\bibitem{mcmahan2017communicationefficient}
B.~McMahan, E.~Moore, D.~Ramage, S.~Hampson, and B.~A. y~Arcas,
  ``Communication-efficient learning of deep networks from decentralized
  data,'' in {\em 20th International Conference on Artificial Intelligence and
  Statistics}, 2017.

\bibitem{li2020federated}
T.~Li, A.~K. Sahu, M.~Zaheer, M.~Sanjabi, A.~Talwalkar, and V.~Smith,
  ``Federated optimization in heterogeneous networks,'' in {\em Proceedings of
  Machine Learning and Systems}, vol.~2, pp.~429--450, 2020.

\bibitem{yao2021fedhm}
D.~Yao, W.~Pan, Y.~Wan, H.~Jin, and L.~Sun, ``Fedhm: Efficient federated
  learning for heterogeneous models via low-rank factorization,'' {\em
  arXiv:2111.14655}, 2021.

\bibitem{qiu2022zerofl}
X.~Qiu, J.~Fernandez-Marques, P.~P. Gusmao, Y.~Gao, T.~Parcollet, and N.~D.
  Lane, ``Zerofl: Efficient on-device training for federated learning with
  local sparsity,'' in {\em International Conference on Learning
  Representations (ICLR)}, 2022.

\bibitem{ye2023heterogeneous}
M.~Ye, X.~Fang, B.~Du, P.~C. Yuen, and D.~Tao, ``Heterogeneous federated
  learning: State-of-the-art and research challenges,'' {\em arXiv:2307.10616},
  2023.

\bibitem{tran2019federated}
N.~H. Tran, W.~Bao, A.~Zomaya, M.~N. Nguyen, and C.~S. Hong, ``Federated
  learning over wireless networks: Optimization model design and analysis,'' in
  {\em IEEE Conference on Computer Communications}, 2019.

\bibitem{caldas1812leaf}
S.~Caldas, P.~Wu, T.~Li, J.~Konecn{\`y}, H.~McMahan, V.~Smith, and
  A.~Talwalkar, ``Leaf: A benchmark for federated settings,'' {\em
  arXiv:1812.01097}, 2019.

\bibitem{9774739}
Z.~Pan and P.~Mishra, ``Hardware acceleration of explainable machine
  learning,'' in {\em 2022 Design, Automation and Test in Europe Conference and
  Exhibition (DATE)}, pp.~1127--1130, 2022.

\bibitem{yang_procrustes_2020}
D.~Yang, A.~Ghasemazar, X.~Ren, M.~Golub, G.~Lemieux, and M.~Lis, ``Procrustes:
  a {Dataflow} and {Accelerator} for {Sparse} {Deep} {Neural} {Network}
  {Training},'' in {\em 2020 53rd {Annual} {IEEE}/{ACM} {International}
  {Symposium} on {Microarchitecture} ({MICRO})}, pp.~711--724, Oct. 2020.

\bibitem{awad_fpraker_2021}
O.~M. Awad, M.~Mahmoud, I.~Edo, A.~H. Zadeh, C.~Bannon, A.~Jayarajan,
  G.~Pekhimenko, and A.~Moshovos, ``{FPRaker}: {A} {Processing} {Element} {For}
  {Accelerating} {Neural} {Network} {Training},'' in {\em {MICRO}-54: 54th
  {Annual} {IEEE}/{ACM} {International} {Symposium} on {Microarchitecture}},
  {MICRO} '21, (New York, NY, USA), pp.~857--869, Association for Computing
  Machinery, Oct. 2021.

\bibitem{qin_sigma_2020}
E.~Qin, A.~Samajdar, H.~Kwon, V.~Nadella, S.~Srinivasan, D.~Das, B.~Kaul, and
  T.~Krishna, ``{SIGMA}: {A} {Sparse} and {Irregular} {GEMM} {Accelerator} with
  {Flexible} {Interconnects} for {DNN} {Training},'' in {\em 2020 {IEEE}
  {International} {Symposium} on {High} {Performance} {Computer} {Architecture}
  ({HPCA})}, pp.~58--70, Feb. 2020.

\bibitem{nakahara_fpga-based_2019}
H.~Nakahara, Y.~Sada, M.~Shimoda, K.~Sayama, A.~Jinguji, and S.~Sato,
  ``{FPGA}-{Based} {Training} {Accelerator} {Utilizing} {Sparseness} of
  {Convolutional} {Neural} {Network},'' in {\em 2019 29th {International}
  {Conference} on {Field} {Programmable} {Logic} and {Applications} ({FPL})},
  pp.~180--186, Sept. 2019.

\bibitem{Lee:ISSCL:2019:mixedp}
J.~Lee, J.~Lee, D.~Han, J.~Lee, G.~Park, and H.-J. Yoo, ``An energy-efficient
  sparse deep-neural-network learning accelerator with fine-grained mixed
  precision of fp8–fp16,'' {\em IEEE Solid-State Circuits Letters}, vol.~2,
  no.~11, pp.~232--235, 2019.

\bibitem{choi_energy-efficient_2020}
S.~Choi, J.~Sim, M.~Kang, Y.~Choi, H.~Kim, and L.-S. Kim, ``An
  {Energy}-{Efficient} {Deep} {Convolutional} {Neural} {Network} {Training}
  {Accelerator} for {In} {Situ} {Personalization} on {Smart} {Devices},'' {\em
  IEEE Journal of Solid-State Circuits}, vol.~55, pp.~2691--2702, Oct. 2020.
\newblock Conference Name: IEEE Journal of Solid-State Circuits.

\bibitem{venkataramani_rapid_2021}
S.~Venkataramani, V.~Srinivasan, {\em et~al.}, ``{RaPiD}: {AI} {Accelerator}
  for {Ultra}-low {Precision} {Training} and {Inference},'' in {\em 2021
  {ACM}/{IEEE} 48th {Annual} {International} {Symposium} on {Computer}
  {Architecture} ({ISCA})}, pp.~153--166, June 2021.

\bibitem{he_enabling_2021}
X.~He, J.~Liu, Z.~Xie, H.~Chen, G.~Chen, W.~Zhang, and D.~Li, ``Enabling
  energy-efficient {DNN} training on hybrid {GPU}-{FPGA} accelerators,'' in
  {\em Proceedings of the {ACM} {International} {Conference} on
  {Supercomputing}}, {ICS} '21, (New York, NY, USA), pp.~227--241, Association
  for Computing Machinery, June 2021.

\bibitem{imani_floatpim_2019}
M.~Imani, S.~Gupta, Y.~Kim, and T.~Rosing, ``{FloatPIM}: in-memory acceleration
  of deep neural network training with high precision,'' in {\em Proceedings of
  the 46th {International} {Symposium} on {Computer} {Architecture}}, {ISCA}
  '19, (New York, NY, USA), pp.~802--815, Association for Computing Machinery,
  June 2019.

\bibitem{jiang_two-way_2020}
H.~Jiang, S.~Huang, X.~Peng, J.-W. Su, Y.-C. Chou, W.-H. Huang, T.-W. Liu,
  R.~Liu, M.-F. Chang, and S.~Yu, ``A {Two}-way {SRAM} {Array} based
  {Accelerator} for {Deep} {Neural} {Network} {On}-chip {Training},'' in {\em
  2020 57th {ACM}/{IEEE} {Design} {Automation} {Conference} ({DAC})}, pp.~1--6,
  July 2020.

\bibitem{esmaeilzadeh2023performance}
H.~Esmaeilzadeh, S.~Ghodrati, A.~B. Kahng, S.~Kinzer, S.~D. Manasi, S.~S.
  Sapatnekar, and Z.~Wang, ``Performance analysis of dnn inference/training
  with convolution and non-convolution operations,'' {\em arXiv:2306.16767},
  2023.

\bibitem{yang2020fpga}
Z.~Yang, S.~Hu, and K.~Chen, ``Fpga-based hardware accelerator of homomorphic
  encryption for efficient federated learning,'' {\em arXiv:2007.10560}, 2020.

\bibitem{saadat2018minimally}
H.~Saadat, H.~Bokhari, and S.~Parameswaran, ``Minimally biased multipliers for
  approximate integer and floating-point multiplication,'' {\em IEEE
  Transactions on Computer-Aided Design of Integrated Circuits and Systems},
  vol.~37, no.~11, 2018.

\bibitem{chen2021pam}
C.~Chen, W.~Qian, M.~Imani, X.~Yin, and C.~Zhuo, ``Pam: A
  piecewise-linearly-approximated floating-point multiplier with unbiasedness
  and configurability,'' {\em IEEE Transactions on Computers}, vol.~71, no.~10,
  pp.~2473--2486, 2021.

\bibitem{li2020accuracy}
J.~Li, Y.~Guo, and S.~Kimura, ``Accuracy-configurable low-power approximate
  floating-point multiplier based on mantissa bit segmentation,'' in {\em 2020
  IEEE Region 10 Conference (TENCON)}, pp.~1311--1316, IEEE, 2020.

\bibitem{Chhajed:CSSP:2022:bitmac}
H.~Chhajed, G.~Raut, N.~Dhakad, S.~Vishwakarma, and S.~K. Vishvakarma,
  ``Bitmac: Bit-serial computation-based efficient multiply-accumulate unit for
  dnn accelerator,'' {\em Circuits, Systems, and Signal Processing}, pp.~1--16,
  2022.

\bibitem{lee_overview_2021}
J.~Lee and H.-J. Yoo, ``An {Overview} of {Energy}-{Efficient} {Hardware}
  {Accelerators} for {On}-{Device} {Deep}-{Neural}-{Network} {Training},'' {\em
  IEEE Open Journal of the Solid-State Circuits Society}, vol.~1, pp.~115--128,
  2021.

\bibitem{cacti}
N.~Muralimanohar, R.~Balasubramonian, and N.~Jouppi, ``Optimizing nuca
  organizations and wiring alternatives for large caches with cacti 6.0,'' in
  {\em 40th Annual IEEE/ACM International Symposium on Microarchitecture (MICRO
  2007)}, 2007.

\bibitem{krizhevsky2009learning}
A.~Krizhevsky, ``Learning multiple layers of features from tiny images,'' {\em
  Master's thesis, University of Toronto}, 2009.

\bibitem{le2015tiny}
Y.~Le and X.~S. Yang, ``Tiny imagenet visual recognition challenge,'' tech.
  rep., Stanford Computer Vision Lab, 2015.

\bibitem{he2015deep}
K.~He, X.~Zhang, S.~Ren, and J.~Sun, ``Deep residual learning for image
  recognition,'' in {\em Proceedings of the IEEE conference on computer vision
  and pattern recognition}, pp.~770--778, 2016.

\bibitem{gong2022approxtrain}
J.~Gong, H.~Saadat, H.~Gamaarachchi, H.~Javaid, X.~S. Hu, and S.~Parameswaran,
  ``Approxtrain: Fast simulation of approximate multipliers for dnn training
  and inference,'' {\em IEEE Transactions on Computer-Aided Design of
  Integrated Circuits and Systems}, pp.~1--1, 2023.

\bibitem{tensorflow2015}
M.~Abadi, A.~Agarwal, P.~Barham, E.~Brevdo, Z.~Chen, C.~Citro, G.~S. Corrado,
  A.~Davis, J.~Dean, M.~Devin, S.~Ghemawat, I.~Goodfellow, A.~Harp, G.~Irving,
  M.~Isard, Y.~Jia, R.~Jozefowicz, L.~Kaiser, M.~Kudlur, J.~Levenberg,
  D.~Man\'{e}, R.~Monga, S.~Moore, D.~Murray, C.~Olah, M.~Schuster, J.~Shlens,
  B.~Steiner, I.~Sutskever, K.~Talwar, P.~Tucker, V.~Vanhoucke, V.~Vasudevan,
  F.~Vi\'{e}gas, O.~Vinyals, P.~Warden, M.~Wattenberg, M.~Wicke, Y.~Yu, and
  X.~Zheng, ``{TensorFlow}: Large-scale machine learning on heterogeneous
  systems,'' 2015.
\newblock Software available from tensorflow.org.

\bibitem{hsu2019measuring}
T.-M.~H. Hsu, H.~Qi, and M.~Brown, ``Measuring the effects of non-identical
  data distribution for federated visual classification,'' {\em
  arXiv:1909.06335}, 2019.

\bibitem{9730377}
Y.~Hu, Y.~Liu, and Z.~Liu, ``A survey on convolutional neural network
  accelerators: Gpu, fpga and asic,'' in {\em 2022 14th International
  Conference on Computer Research and Development (ICCRD)}, pp.~100--107, 2022.

\bibitem{mitchell1962computer}
J.~N. Mitchell, ``Computer multiplication and division using binary
  logarithms,'' {\em IRE Transactions on Electronic Computers}, no.~4, 1962.

\bibitem{10000632}
M.~Kim, O.~Günlü, and R.~F. Schaefer, ``Effects of quantization on federated
  learning with local differential privacy,'' in {\em GLOBECOM 2022 - 2022 IEEE
  Global Communications Conference}, pp.~921--926, 2022.

\bibitem{kang2024effect}
T.~Kang, L.~Liu, H.~He, J.~Zhang, S.~Song, and K.~B. Letaief, ``The effect of
  quantization in federated learning: Ar$\backslash$'enyi differential privacy
  perspective,'' {\em arXiv:2405.10096}, 2024.

\bibitem{wei2020federated}
K.~Wei, J.~Li, M.~Ding, C.~Ma, H.~H. Yang, F.~Farokhi, S.~Jin, T.~Q. Quek, and
  H.~V. Poor, ``Federated learning with differential privacy: Algorithms and
  performance analysis,'' {\em IEEE transactions on information forensics and
  security}, vol.~15, pp.~3454--3469, 2020.

\end{thebibliography}

\end{document}